# Building and Managing a Tropical Fish Facility: A Do-It-Yourself Guide

Claudius F. Kratochwil[1,2,*] , Muktai Kuwalekar[1,2,#] , Jan Häge[1,2,#] , Nidal Karagic[1,2,#]

[1] Helsinki Institute of Life Science (HiLIFE), University of Helsinki, Finland
[2] Faculty of Biological and Environmental Sciences, University of Helsinki, Finland

* Corresponding author: Claudius.Kratochwil@helsinki.fi
# These authors contributed equally

**At the core of most research in zoological disciplines, ranging from developmental biology to genetics to behavioral biology, is the ability to keep animals in captivity. While facilities for traditional model organisms often benefit from well-established designs, construction of a facility for less commonly studied organisms can present a challenge. Here, we detail the process of designing, constructing, and operating a specialized 10,000-liter aquatic facility dedicated to housing cichlid fishes for research purposes. The facility, comprising 42 aquaria capable of division into up to 126 compartments, a flow-through rack for juveniles, egg tumblers for eggs and embryos, and a microinjection setup, provides a comprehensive environment for all life stages of cichlid fishes. We anticipate that a similar design can be also used also for other tropical teleost fishes. This resource is designed to promote increased efficiency and success in cichlid fish breeding and research, thereby offering significant insights for aquatic research labs seeking to build or optimize their own infrastructures.**

Aquaculture | Fish Facility | Aquatic animals | Cichlid Fishes | #NewPI

## 1 | INTRODUCTION

Cichlid fishes represent a highly diverse group which have been the subject of extensive interest in the fields of evolutionary biology, ecology and more recently genetics and developmental biology. Their elaborate behavioral traits, rapid rate of speciation, and varied morphology make them ideal subjects for scientific research (Santos, Lopes, and Kratochwil 2023) as well as outreach (Magalhaes and Ford 2022). However, maintaining these fishes and conducting research under controlled conditions necessitates specific aquatic housing solutions that cater to their unique needs. While there is an abundance of excellent literature on constructing facilities for zebrafish (McNabb et al. 2012; Bhargava 2018; Paige et al. 2014), we did not find any comprehensive hands-on guide for building a facility intended for larger (compared to zebrafish) tropical freshwater fish as cichlid fishes. Therefore, we have taken the initiative to provide such a guide. This guide will first cover general considerations for building a fish facility, which may be useful for all researchers planning to construct a mid-sized facility, and then commence with more specific aspects of facility construction using our own 10,000-liter facility as an example. The aim of this manuscript is to provide insight into the key aspects to consider when constructing such a facility and to offer guidance throughout the process. We hope that this guide will not only facilitate and motivate researchers but also encourage them to explore a broader range of organisms for their research.

## 2 | GENERAL CONSIDERATIONS

When embarking on the construction of a fish facility (**Figure 1**), there are several crucial and far-reaching decisions that need to be made. These decisions play a significant role in determining the success and long-term sustainability of the facility. It is understandable that these decisions can be slightly stressful, as they often represent the or one of the biggest investments for a young research group. However, it is important to remember that with careful planning and thoughtful consideration, these challenges can be overcome.

During this process, open communication with colleagues and experts in the field can be incredibly helpful. Engaging in discussions with peers who have experience in building fish facilities can provide valuable insights into the potential challenges and reassure you that, with proper planning, the project will succeed in the end. Their advice and shared experiences can help identify potential pitfalls, offer solutions, and provide the necessary support and trust that the facility will be running smoothly. By approaching the decision-making process with a combination of careful consideration, expert guidance, and open communication, one can greatly improve the feasibility of your undertaking as well as the reliability and usability of the final facility. Very importantly, this will also give you a chance to alleviate some of the associated stress and potentially find ways to increase the capability of the facility beyond the initial scope.

- **Space Availability:** Evaluate the available space for the fish facility. Determine the size of the room or area and consider whether and how it can accommodate the necessary equipment, fish tanks, and operational workflow.

- **Funding:** Evaluate the financial resources available for both the initial construction of the facility and its ongoing maintenance. Determine the budget required for equipment, tanks, filtration systems, utilities, fish food, and any other necessary expenses. It is important to note that there is often a significant cost difference, which can range from a factor of 5 to 10-fold, between commercial solutions and self-built alternatives. Consider the advantages and disadvantages of each option based on your budgetary constraints, know-how, time availability and long-term sustainability goals. It may be reasonable to set aside apart a not to small part of the budget to be able to handle the replacement of faulty equipment or other unplanned expenses without delay.

- **Fish Species and Requirements:** Determine the specific fish species you intend to house in the facility. Research their individual needs, such as water temperature, pH levels, water hardness, oxygen requirements, substrate needs, and tank size. Ensure that you can provide the optimal environment for the chosen fish species.

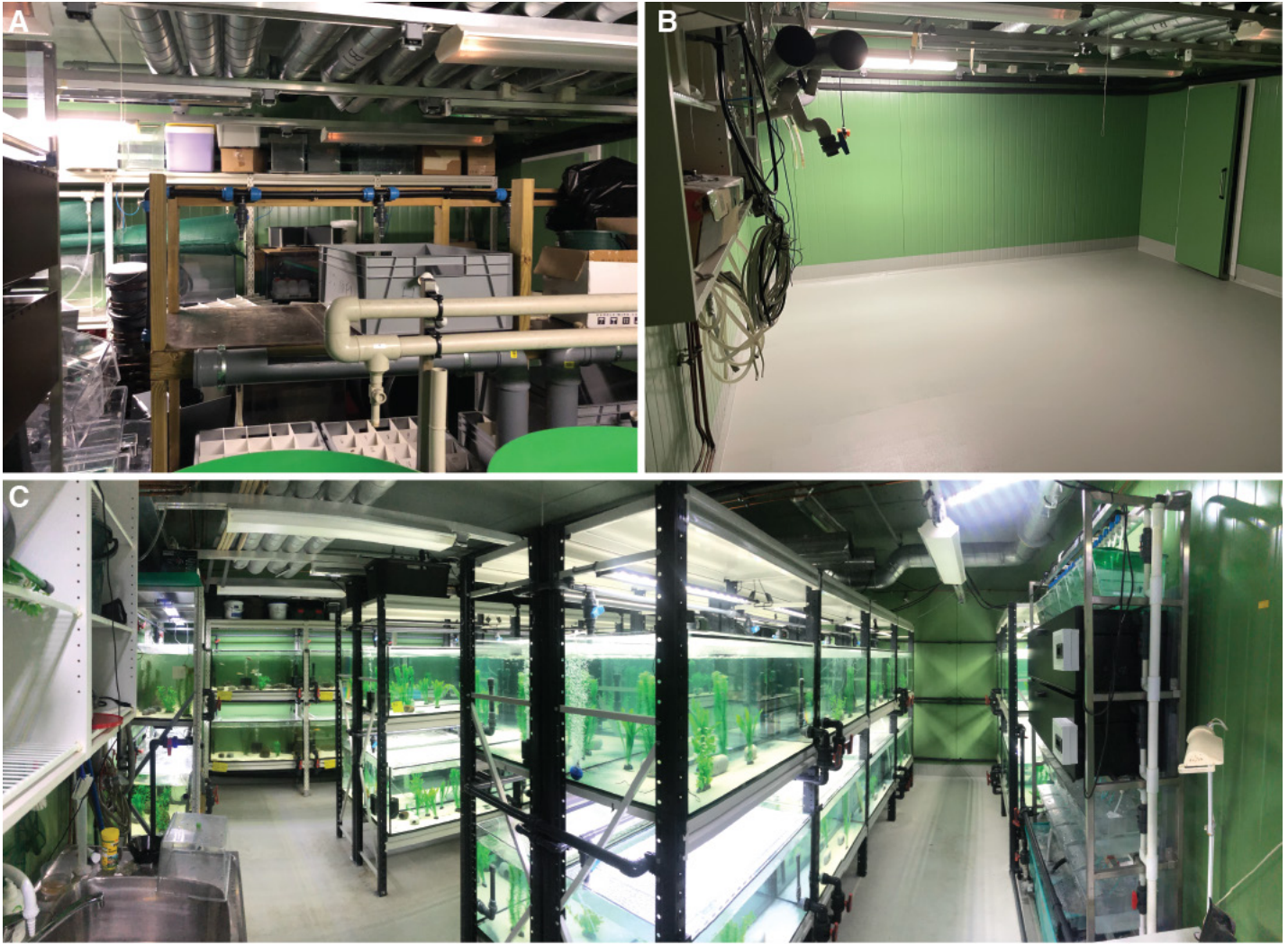

**Figure 1 | The process of building a facility.** Building a facility from scratch can be a challenging task, especially for a young research team. **(A)** It starts with the decision of what to keep from former users of the room and getting everything cleaned up. **(B)** However, the challenges do not diminish with an empty room, as one must make long-term decisions about how the room will be configured. **(C)** Establishing the facility is a milestone, but it is only the beginning of the next journey, which is to maintain a functioning and organized facility and a productive working environment capable of addressing unforeseen challenges.

- **Timeline:** Determine your timeline for the fish facility's partial or full operation. Consider the option of building in stages to meet urgent research needs first and gradually expand the facility. Phased construction allows for effective financial management, flexibility in design and functionalities, and the opportunity to refine and optimize future expansions based on lessons learned from the initial phase.

- **Research Objectives:** Clarify your research goals and determine the number of fish required in the short and long term. Consider factors such as the frequency of experiments, breeding requirements, and sample sizes needed for your research.

- **Collaboration and Decision Making:** Engage in discussions with relevant people within your university to determine who needs to be involved in the decision-making process regarding the fish facility. Deans, directors, veterinarians, university architects, or electricians may have valuable input, potential concerns, or disagreements that could influence your plans. Seek their expertise and address any potential issues early on to ensure a smooth implementation.

- **Additional Room Requirements:** Consider the need for additional space within the room, beyond the fish tanks. Depending on your research objectives, it may be advantageous to allocate space for equipment such as microinjection setups, microscopes, computers for data recording, worktables, video or photography setups, or 2D shakers for raising embryos. Assess the requirements of your specific research projects and plan the room layout accordingly.

- **Heating:** Evaluate the heating situation of the room. If you plan to house tropical fish species, maintaining warm temperatures is crucial. Determine if the room is already heated or if additional heating measures are required. Consider whether precise temperature control is necessary or if a certain temperature range is acceptable. Additionally, decide whether you prefer to heat individual tanks using heaters or if alternative heating methods are available and suitable for your needs.

- **Other External Factors:** It is important to consider various external factors that can impact its functionality. If the room has windows, managing light-dark rhythms and controlling algae growth under direct sunlight can present challenges. Ideally, the facility would be sheltered from external light and illuminated with broad spectrum LED lamps that operate automatically in a natural light-dark-cycle. Ineffective ventilation can also affect the room's humidity due to the presence of a large water surface. To address these issues, it may be necessary to incorporate a dehumidifier into the facility. This helps maintain optimal humidity levels, ensuring a better working environment, preventing mold development, and safeguarding the proper functioning of technical equipment, especially electronic and optical devices.

- **Water Supply:** Assess the availability and quality of the water supply in the room or building. Consider factors such as water source, water pressure, water temperature, water hardness, pH levels, and any potential contaminants as chlorine or dissolved metals that may be harmful to the fish. Consider water treatment systems if necessary. Sufficient water pressure is particularly important to assess for some aquarium systems, including the one described here. In case the water pressure is not sufficient, a pumping system and an elevated reservoir might be needed.

- **Water Change Frequency:** Decide on the desired frequency of water changes based on the fish species and number, facility size, and available resources. Water changes can range from constant change (flow-through systems) to regular exchanges in large volume sump systems or manual changes in individual aquaria. Consider the impact on fish wellbeing and requirements, start-up funding, and the time you can invest in maintenance.

- **Drainage System:** Evaluate the facility's drainage system thoroughly to ensure effective water management and prevent potential flooding. Adequate drainage is vital for maintaining a healthy environment. The drainage rate will also play a role in determining the organization of water changes and the consideration of implementing a flow-through system for efficient water circulation.

- **Experiment Types:** Determine the nature of the experiments you plan to conduct in the facility. Consider whether you will focus solely on maintenance or if you will also conduct behavioral experiments, breeding, or genetic crossing. This will influence the size and type of tanks required.

- **Labeling system.** In a facility that accommodates numerous users, species, and strains, an efficient labeling system is essential. The ideal system should allow for easy application, repositioning, and removal of labels while minimizing the risk of loss of information. Additionally, having the capability to label both the quantity and type of food isadvantageous. Electronic systems offer the benefit of data storage, but they lead to a significant increase in the time needed for data entry.

- **Disease Prevention:** Carefully consider from where to buy fish and evaluate the need to regularly introduce new fish into the facility. Consider the potential risks associated with disease transmission and implement appropriate quarantine protocols to minimize the introduction of pathogens or contaminants. Are there any particularly troublesome fish diseases for your species (e.g., Tilapia Lake Virus for cichlid fishes) and in what regions are they found? Unfortunately, for many tropical fish species little information is available regarding specific disease susceptibility. It can make sense to use the more extensive literature on related commercially used species. Major diseases of relevance to cichlids can for example be found in the Nile Tilapia fact sheet from the Cultured Aquatic Species Information Program of the Food and Agriculture Organization of the United Nations (FAO) (Rakocy 2006).

- **Power Supply and Utilities:** Assess the availability of power plugs in the facility for lighting, heaters, and other equipment. Especially during water change with cold water, heaters need a significant amount of power that can lead to circuit overloads if the heaters are not distributed over different circuits. Additionally, consider whether compressed air piping is necessary for aeration and sponge filters. Ensure that the facility's utilities can support the operational requirements.

- **Maintenance Responsibilities:** Determine who will be responsible for maintaining the fish facility. Assess whether technical staff will be available or if lab members will take care of the facility and how weekends, holidays and holiday seasons will be handled. Consider the workload required per week and ensure it aligns with the available resources and long-term sustainability of the facility.

- **Consideration of Hazards:** Finally, but very importantly, assess potential hazards to humans, animals, the building, and the environment within the fish facility. Contemplate scenarios such as broken aquaria, tubing failures, fires, natural disasters, power outages, compressed air system malfunctions, or technical failures in any operational systems within the room. Evaluate the effects of both shorter and longer durations of these incidents. Implement safety measures and develop contingency plans to mitigate risks and minimize the impact of such events on the well-being of the fish, the facility, and the surrounding environment. This for example also includes getting conformation on proper sealing of the floor, as floor renovations will be challenging once the facility is in place.

These initial considerations will help guide the design, construction, and operation of your fish facility, setting the stage for a successful and productive research environment. It is also recommended to consult with experts or experienced researchers, to ensure all necessary aspects are considered.

## 3 | THINKING IT THROUGH: A CASE STUDY OF A CICHLID FISH FACILITY

In this chapter, we share our thought process of designing a specialized facility for our Integrative Evolutionary Biology lab at the University of Helsinki, Finland. In our research we focus on the Genetics, Genomics, and Evolutionary Developmental Biology of East African cichlids. Our primary objectives for our research were to create a space that would enable us to maintain 10-15 species, to perform hybrid crosses and raise $F_1$ and $F_2$ generations in sufficient number and to independently raise clutches of embryos for developmental analyses. These considerations heavily influenced our design choices. To do so we had to create an optimal research environment aligned with our objectives and needs. We encountered various considerations that we already discussed in the second chapter, from space availability and funding to long-term maintenance plans and potential hazards. Collaboration and open communication with colleagues and experts played a vital role in guiding our decisions.

Our personal journey should serve as a relatable account, offering practical guidance for fellow researchers. By sharing what worked for us and what lessons we learned, we hope to assist others in making informed choices and navigating the complexities of building their own specialized research facilities.

The decision about the general setup necessitates consideration of several interdependent factors. Most importantly among these is the availability of physical space, which shapes the facility's design and capacity, influencing tank numbers, equipment layout, and workflow design. Moreover, funding constraints, both for building and running costs, might impact the construction, maintenance, and operation of the facility, as well as personal involvement in its construction and maintenance. Accommodating the distinct requirements of fish species, such as water quality parameters, is crucial for their well-being and successful breeding. Furthermore, the facility's design must align with the needs of various research experiments, necessitating adaptable configurations to cater to tanks for raising, maintaining, isolating, and experimental setups that eventually include specifically controlled environments. Throughout the entire process, there must be communication with necessary authorities and involved parties, whether it be veterinarians, ethics boards, university architects, neighboring labs and facilities, electricians, directors, or deans.

**Space and Funding.** In our case, we had access to ca. 32m$^2$ (ca. 5.3 x 6m) of space to construct our fish facility and related infrastructure. Our primary goal was to maximize the available space for housing the fish, while still allowing comfortable movement within the facility. Additionally, we allocated an area for a table equipped with a microinjector and a microscope to perform microinjections, ensuring we had the necessary tools for our research activities. The inclusion of a sink in the room was a nice addition as it is essential for cleaning purposes of filters and other equipment. At the beginning it is important to make a floor plan to plan precisely where aquaria and other equipment will go. Funding-wise we luckily had access to flexible start-up funding, which allowed us to dynamically set priorities for lab equipment and fish facility. After exploring commercial solutions that would have cost between at least 150,000€ for a very limited aquarium space, we opted for a custom solution that we built ourselves, reducing the cost to roughly 30,000€ while being much more functional and spacious. This decision carried the risk of potential challenges and failures, but it provided us with the opportunity to build the facility to meet our specific needs.

**Fish species and their space, water, and substrate requirements.** Our goal was to accommodate around 10-15 species of East African Haplochromine cichlids from Lake Malawi and Lake Victoria. These fish require relatively large aquaria, ideally exceeding 150 liters. Among the critical environmental considerations for these fish, water hardness took precedence. In Lake Malawi, water hardness can range from 10 to 20 degrees of hardness (dH), which significantly surpasses the average tap water hardness in Finland, usually ranging from 2 to 4 dH. Consequently, we needed to manipulate water parameters to establish a suitable environment for our cichlids. This simultaneously ruled out the possibility of employing a flow-through system (as salt would be washed out) and using smaller aquaria (owing to difficulties in maintaining stable hardness levels). Additionally, the water temperature couldn't be controlled.

To enhance water hardness, we introduce buffering substances (Sodium hydrogen carbonate $NaHCO_3$ and Magnesium Sulphate, $MgSO_4 \times 7H_2O$). This helped stabilize the pH at 8.0 and elevate hardness levels (GH 7, KH 10) within the desired range for East African cichlids (pH: 7.6 - 8.4; GH (General Hardness): 4 - 12 dGH; KH (Carbonate Hardness): 5 - 15 dKH). As salt would need to be added during each water change, this influenced our decision to employ larger aquaria, as a greater water volume fosters more stable parameters. The considerations pertaining to water hardness also guided our selection of a substrate—Bahamian Oolitic Aragonite Sand. This substrate, primarily composed of Oolitic Aragonite, is directly sourced from the Bahamas and is free from additives. Its value lies in its high calcium carbonate content, which helps maintain optimal water hardness levels. It also boasts excellent buffering capabilities, thereby minimizing pH fluctuations. The sand is also conducive to beneficial bacteria and is both easy to clean and maintain.

**Choosing aquaria and shelves.** The chosen aquarium type (**Figure 2**) for the facility comprises a uniform volume of 230 liters. Each individual aquarium is characterized by dimensions measuring 99 cm in length, 50 cm in width, and 47 cm in height. The construction of these aquaria involves the utilization of glass panels, notably 8 mm in thickness, which are adjoined using black silicon to ensure robust sealing and structural cohesion. The incorporation of stress bars serves to augment the aquaria's overall stability and longevity. Moreover, these stress bars facilitate the incorporation of lids, a subsequent addition to the aquaria. To accommodate lids and provide supplementary compartmentalization, 5mm thick transparent acrylic sheets, specifically polymethyl methacrylate (PMMA), were employed. While constituting a comparatively more expensive investment, the utilization of PMMA bears distinctive advantages, encompassing its lightweight nature, optical clarity marked by high light transmission properties, compliance with food safety standards, and the possibility to modify the sheets by drilling holes. One of the disadvantages of PMMA is that it can warp slightly, but the sheets can be turned from time to time to make them return to the original shape. One of the most expensive items were the shelves to carry the aquaria. Yet, due to the substantial weight they must support (a single aquarium filled with water weighs nearly 250 kg), careful consideration is necessary. As a solution, we have chosen to procure heavy-duty shelves that offer the capability of being interconnected. These shelves not only provide the necessary strength to accommodate significant loads but also allow for the simultaneous installation of cables, lighting, pipes, and tubes. Additionally, they offer storage space on the top surface.

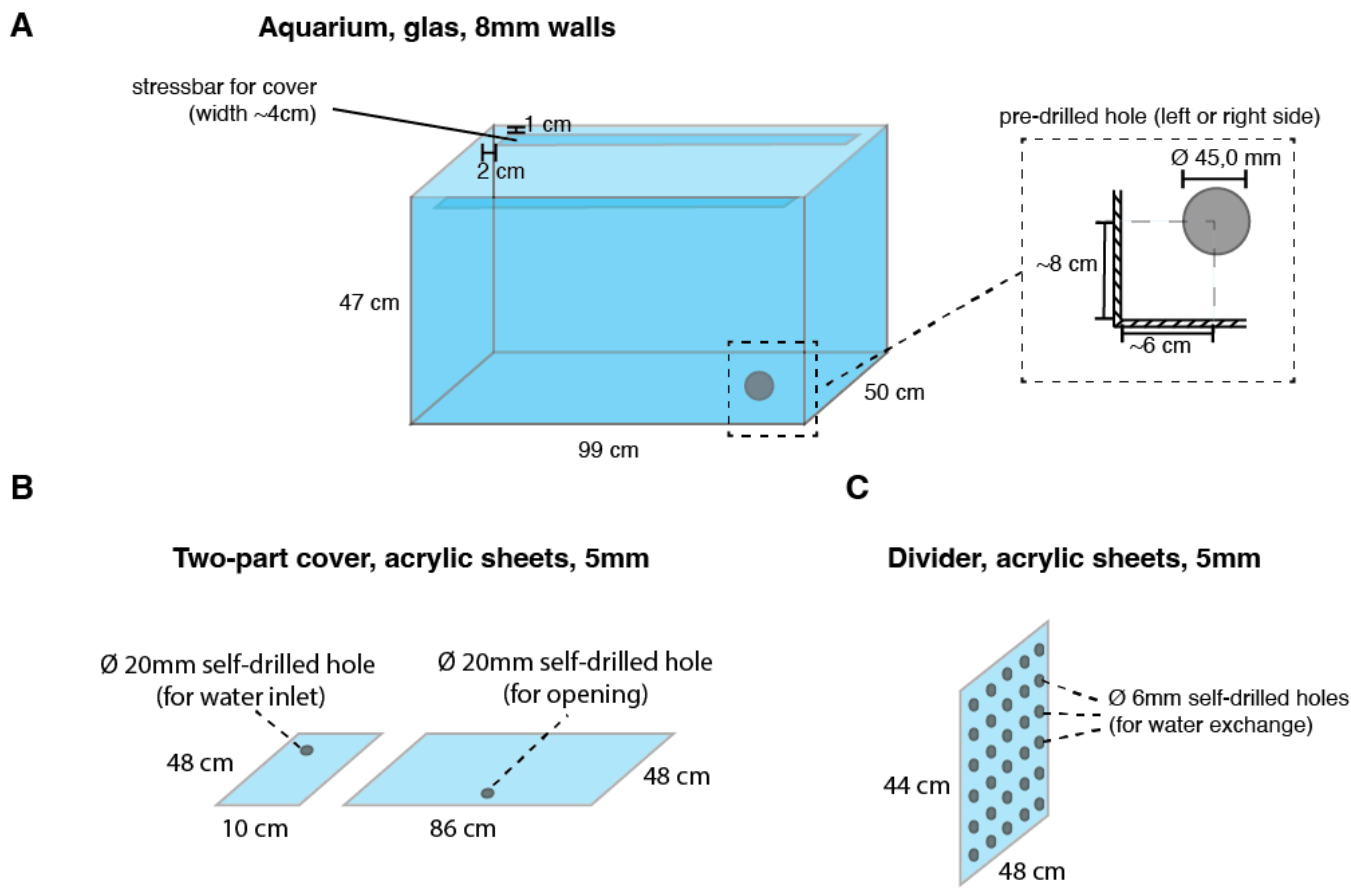

**Figure 2 | Aquarium design. (A)** When ordering aquaria, several decisions need to be made, including the size, wall thickness (8mm is sturdier than the common 6mm used in smaller aquaria), stress bars (also serving as lid supports), and whether pre-drilled holes are required, along with their specific size and placement. We opted for 230-liter aquaria, with dimensions determined by the available shelves. It's important to allocate sufficient space for handling above the aquaria, ensuring that standard buckets can fit through. Hole size and placement were discussed in consultation with the aquarium manufacturer. **(B)** For lids, glass is the more economical choice, but acrylic sheets are easier to drill and less prone to breakage. In our facility, we utilized two-part covers. The smaller piece holds the water inlet securely in place. **(C)** Acrylic sheets also serve as excellent material for dividers. We used a standard 6mm (or 8mm) drill to create holes, facilitating water exchange between compartments.

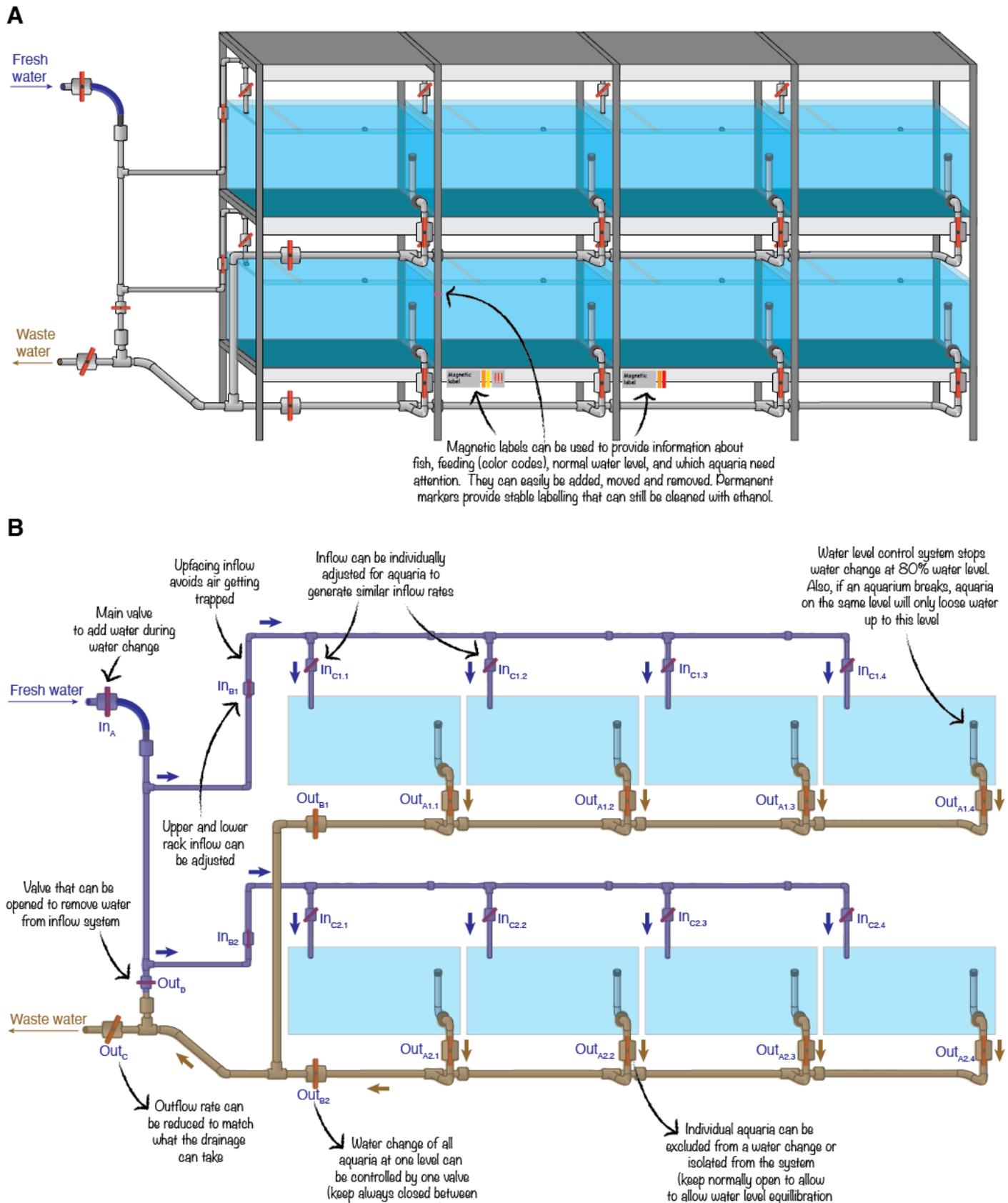

**Figure 3 | Shelf and piping design. (A)** The panel presents a simplified illustration of our aquarium facility featuring four interconnected racks (2m height; 1m width) housing eight 230-liter aquaria. The system is designed to expedite water changes while mitigating the risks associated with broken aquaria and disease. The aquaria are semi-isolated and can also be fully isolated for optimal control. **(B)** A more detailed schematic demonstrates the system's functionality. During the water change process, OutB1 and OutB2 are opened, allowing the removal of 20% of water from all eight aquaria. OutC can adjust the flow rate to the drainage if needed. To introduce new water, InA is opened, and the InB and InC valves are adjusted to ensure uniform water inflow. Minor discrepancies are balanced out as the aquaria are semi-connected. Out D can be utilized at the end to clear the tubes of any remaining water.

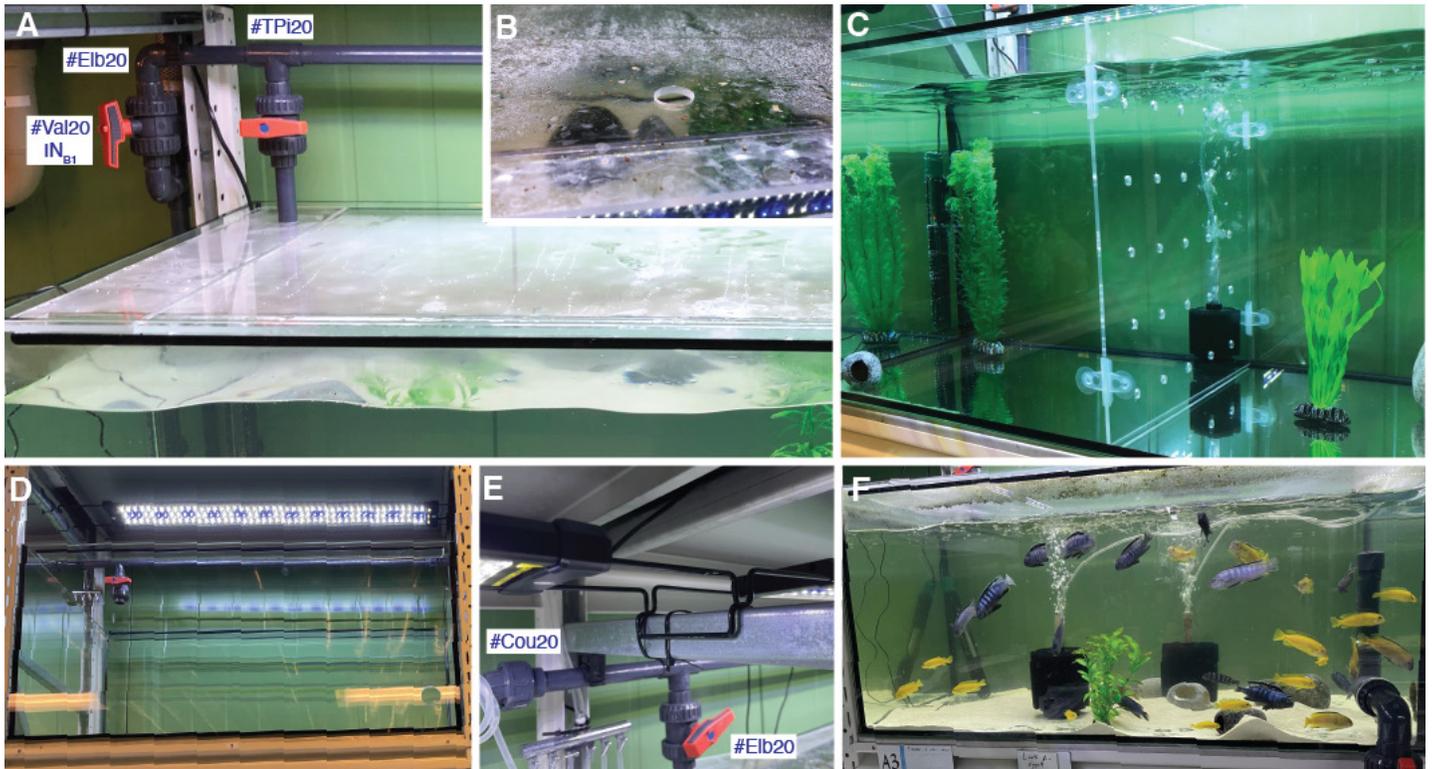

**Figure 4 | Aquaria and their equipment.** **(A)** Acrylic sheet lid with inflow system. **(B)** Self-drilled hole (using 20mm hole saw) in acrylic sheet lid to be able to easily open lids. **(C)** Acrylic sheet divider with self-drilled holes (8mm) and heater, artificial plants, and a slightly smaller sponge filter that we tried at the beginning. **(D)** Lightning system with LED lights over every aquarium. **(E)** Lights are simple fixed with cable ties. **(F)** A fully running aquarium with sponge filters, heater, sand, artificial plants and hiding places.

**Finding creative solutions to reduce maintenance time and costs.** When it comes to maintaining a facility with multiple aquaria, the most time-consuming tasks are water change and cleaning of the aquaria. Our objective was to develop a system that is safe, easy to use, and minimizes risks related to material damage, human error, and disease. We excluded a flow-through system due to issues with water hardness, temperature, and costs. While a sump system was another option, we opted against it due to its complexity, space requirements, and the potential for leaks and overflows. However, we wanted to avoid an isolated aquarium system that is time-intensive to maintain. Thus, we designed a system where aquaria could be simultaneously drained to 80% using a valve (**Figure 3**). This design also allowed for individual aquarium isolation and was failsafe, ensuring that a single aquarium breakage would not lead to complete emptying of the system. Similarly, we aimed for an efficient water inflow system, enabling all aquaria to be refilled using a valve. To enhance the aquaria system, the only alteration required for both the aquarium units and their lids was to place an order for aquarium units that featured a predrilled hole at one of the lower corners to facilitate outflow (**Figure 2A**). We divided the lid into two sections: a smaller, non-opening part which included the hole for water inflow preventing water spillage, and a larger part that can be easily opened without any hindrance (**Figure 4A**).

**Lighting and heating.** The last decisions to be taken at the beginning pertained to heating, lighting, filtration, and aeration. Regarding heating, we opted for standard sera 150W heaters (**Figure 4C, F**). An alternative consideration was room heating; however, this option is not only more costly but also exhibits greater variability. The temperature can be monitored with thermometers. We recommend non-digital options, as batteries of digital options have to be frequently replaced. In terms of lighting, a crucial aspect was the capacity to modulate intensity and on/off timing while minimizing energy consumption. Consequently, we selected NICREW LED lights (**Figure 4D, E**), which boast high efficiency (180 LEDs), low power usage, water resistance, and extendable brackets to accommodate varying aquarium widths. Moreover, these lights offer the flexibility to control the on/off and brightness settings of both the white and blue LEDs either independently or simultaneously.

**Filtration and oxygenation.** The final deliberation encompassed aeration and filtration. The top criteria here were good filtration, easy maintenance, and low risk of outage. As we had a compressed air system installed in the room, we decided to have two large sponge filters per aquarium (**Figure 4C, F**). Sponge filters offer a range of advantages that make them a practical choice for aquarium filtration. They excel in mechanical filtration by effectively trapping larger debris and particles from the water column. Furthermore, their porous structure provides ample surface area for beneficial bacteria, supporting biological filtration and the conversion of harmful ammonia and nitrite into less toxic nitrate. These filters also function as aeration devices, generating gentle water movement and surface agitation to promote oxygen exchange. This makes them suitable for aquaria with delicate or slow-moving fry, as they create a mild current that does not harm them. Additionally, their reliance on compressed air for operation means that they are less affected by power outages, as compressed air systems are often restored early due to their involvement in critical processes. Another benefit lies in their ease of maintenance – a simple rinsing of the sponge during regular water changes suffices for cleaning. However, it is important to note that while sponge filters do have disadvantages as they might not be sufficient for heavily stocked aquaria. Such aquaria might need additional mechanical filtration to prevent the clogging of the sponge material which would compromise biofiltration. Yet, in our experience two large sponge filters (height of sponge material 10cm, diameter 11cm) in combination with our substrate that also contributes to biofiltration and a biweekly filter cleaning regime turned out to be sufficient for a 230-liter aquarium with medium densities of fish. It should be noted that biofiltration efficiency, and thus the amount of internal surface needed in the filtration unit, depends strongly on factors such as water flow, salinity, pH, oxygen and most importantly temperature. Hence our practice as described above might not be applicable under different conditions and especially if lower water temperatures are used. The amount of biofilter media needed given a certain daily feed input can be for example calculated using freely available resources like the FAO handbook on Small-scale aquaponic food production (Carruthers 2015). Very importantly, when new aquaria are used for the first time an initial system cycling has to be performed in order to establish bacteria populations. This process may take 3-5 weeks and needs some regular effort in terms of adding an ammonia source by monitoring the levels of different nitrogen compounds.

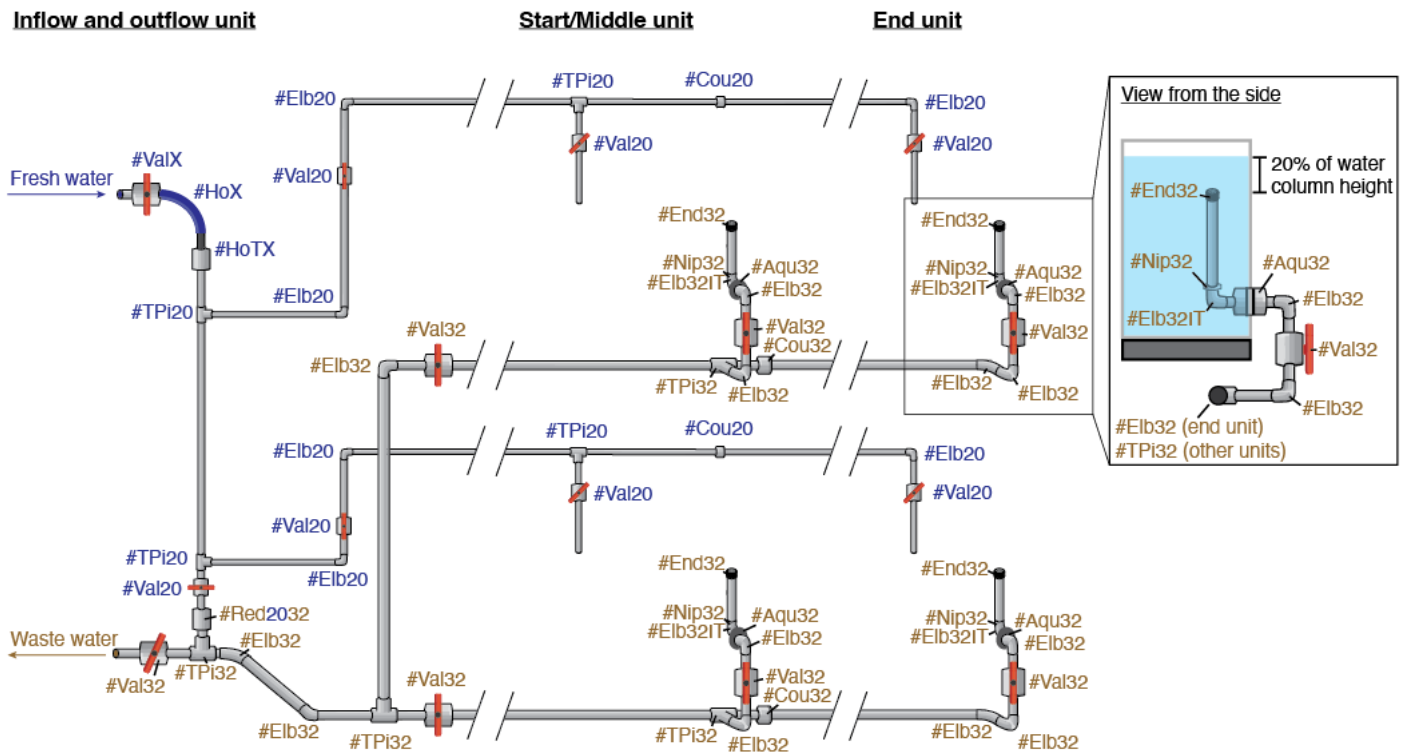

**Figure 5 | List of items and where they are installed.** This figure provides a summary of all items that are built into the facility.

## 4. | MAKING IT HAPPEN: DECISIONS, COSTS AND THE STORY OF BUILDING OUR CICHLID FISH FACILITY

### 4.1 | Deciding on and starting the implementation.

The next step was to start building and testing the facility. We implemented a two-phase construction approach for our fish facility. The first phase involved building a quarter of the facility, which allowed us to begin operations promptly and test the facility on a smaller scale. Interestingly, the first phase did not necessitate significant changes (a few listed below) or optimization to our original plan, indicating the soundness of our initial designs and operations. However, the greatest value derived from this phase was the confirmation that our general approach worked. Having witnessed the functionality and success of our strategy on a smaller scale, it significantly reduced the stress and uncertainty associated with the large investment of constructing the remaining three-quarters of the facility in the second phase. Hence, the phased approach not only got us up and running quickly, but it also gave us the confidence to move forward with the construction of the whole facility.

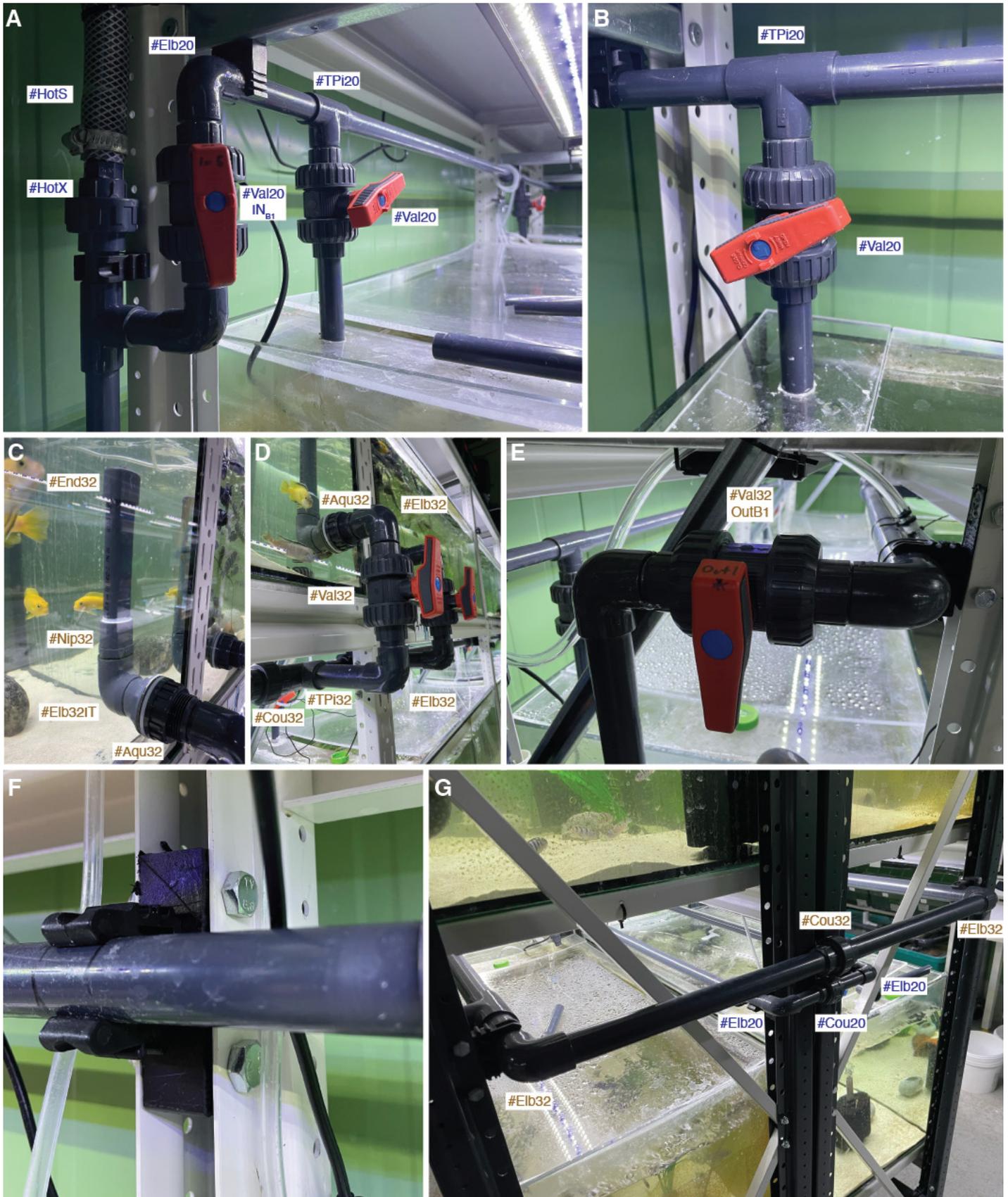

**Figure 6 | Piping construction. (A)** Inflow system of the upper level. **(B)** Inflow at one aquarium. **(C)** Side view on aquarium feed through and water level control system. **(D)** Bottom view on aquarium feed through and water level control system. **(E)** Water change outflow control. **(F)** Pipes can be fixed to the rack using polyethylene blocks and pipe clamps. **(G)** The system can be also extended across racks. The only important thing is that the piping is horizontal. Using couplings in regular intervals eases building the system and allows to replace parts if necessary.

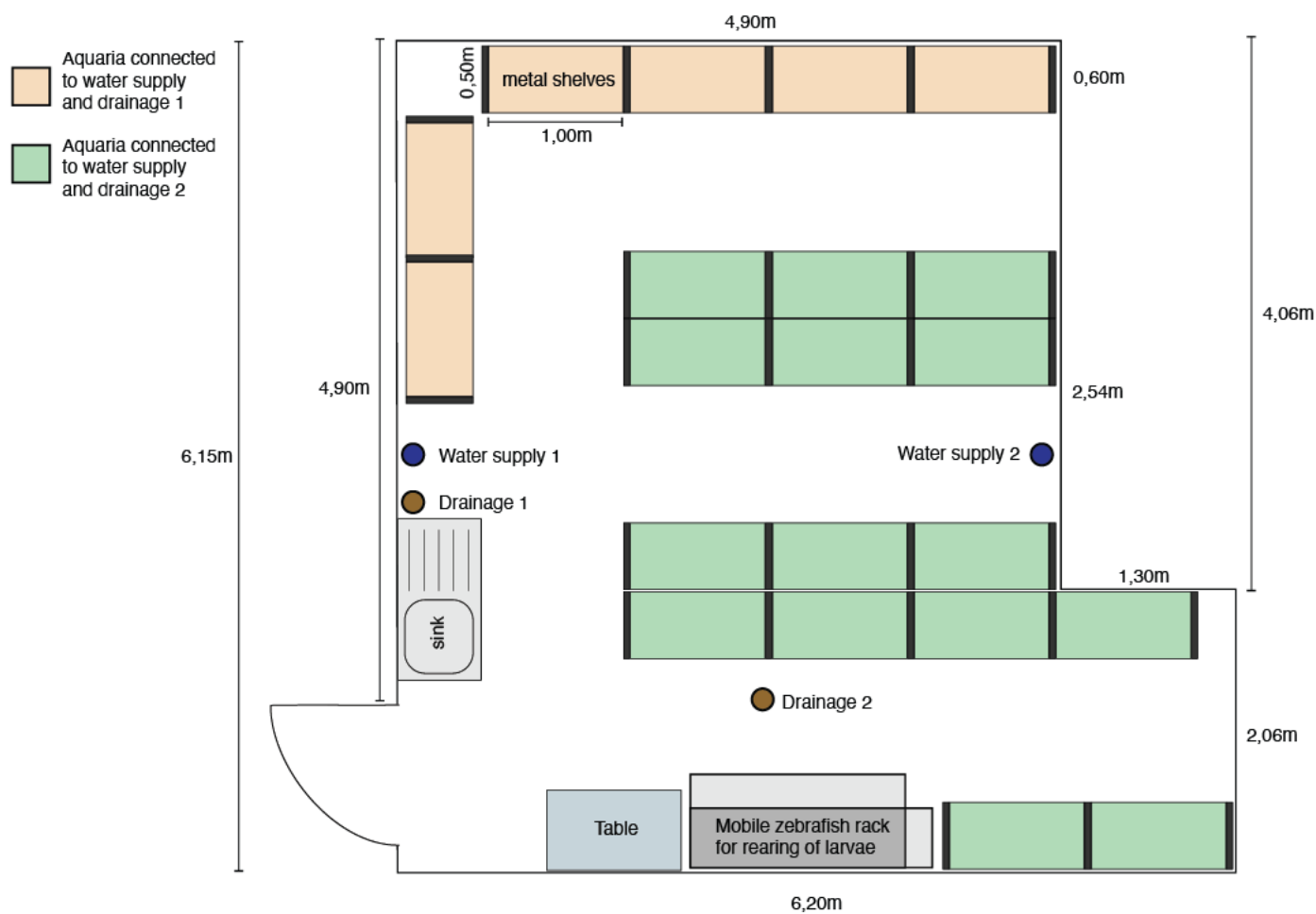

**Figure 7 | The overall designing of a facility.** After designing the individual racks, it becomes crucial to contemplate the most efficient utilization of the available space. Providing sufficient room for additional tasks, equipment, water supply, and drainage connections is of paramount importance. In our case, all aquaria located on the same level within racks of the same color are semi-connected. This implies that water changes across the entire facility are streamlined by merely opening four valves—two from the upper level and two from the lower level of the green and orange systems—to remove 20% of the water. Subsequently, by closing these valves and opening two others (green and orange system), water is replenished in all aquaria simultaneously.

**Installation of the aquaria.** The installation of the facility proceeded surprisingly smoothly. Initially, shelves were set up, and the aquaria were carefully positioned on green camping mats to mitigate the risk of glass fractures from stress. Following this, outflow and inflow pipes were introduced into the setup. For this purpose, we utilized PVC pipes, fittings, and valves (**Figures 5 and 6**). Particularly for valves it is important to ensure a high quality, which means that they do not only effectively hold the water pressure, but that they can also be smoothly opened and closed. To ensure accurate lengths, pipes were cut using a reliable pipe cutter, which saved time and simplified the task. To ensure seamless pipe connections, a deburring tool was employed on both the inside and outside ends of each pipe. This step created smooth, chamfered edges, facilitating even distribution of solvent for effective pipe joining. Before the gluing process, we applied PVC cleaner to eliminate any residues and impurities from areas where glue would be used. The components were then bonded using PVC adhesive, a process known as solvent welding. The elements are then carefully joined together, and during the subsequent five minutes, they must remain undisturbed. This brief waiting period is crucial for the adhesive to establish a strong initial bond between the components. After the five-minute interval, the connected elements are left to dry for a minimum of 24 hours. This extended drying duration is essential to ensure the adhesive undergoes a thorough curing process. Over this time, the adhesive undergoes a chemical reaction that leads to the fusion of the PVC material. As a result, a robust and long-lasting connection is formed. For the cleaning and gluing of the elements, safety instructions should be read. Ensuring proper ventilation is essential, along with using suitable protective gear like gloves and clothing. The setup's overall arrangement is illustrated in **Figures 5 and 6**. Incorporating PVC screw connections with adhesive sockets at intervals simplifies future installation, replacement, and maintenance tasks. For secure placement, pipe clamps are utilized (**Figure 6F**), attaching them to the shelf's vertical bars using wooden or high-density polyethylene blocks. After a minimum of 24 hours to allow the adhesive to set, the system can be tested. Minor leaks, if any, can be resolved by draining the pipes and applying underwater adhesive (e.g., Hobby fix Aquarium glue). This process can be repeated until all leaks are fixed. Following this, all aquaria should be filled and tested to ensure both the tanks and interconnected pipes hold water effectively.

**Equipping the aquaria.** Next on the list is to add filters, lights, heaters, substrate, aquarium enrichment like plastic plants and hiding places, thermometers, and lids. This process should be relatively straightforward. Filters can be connected via valve manifolds and adapters to the compressed air system. Lights are mounted onto the horizontal metal bars, and the light level and on/off cycle are set up. Heaters are simply added to the back of the aquarium. For electric plug it is important that they are placed above the aquaria to avoid that water reaches them. The substrate, plastic plants, and hiding places are then added to the aquarium.

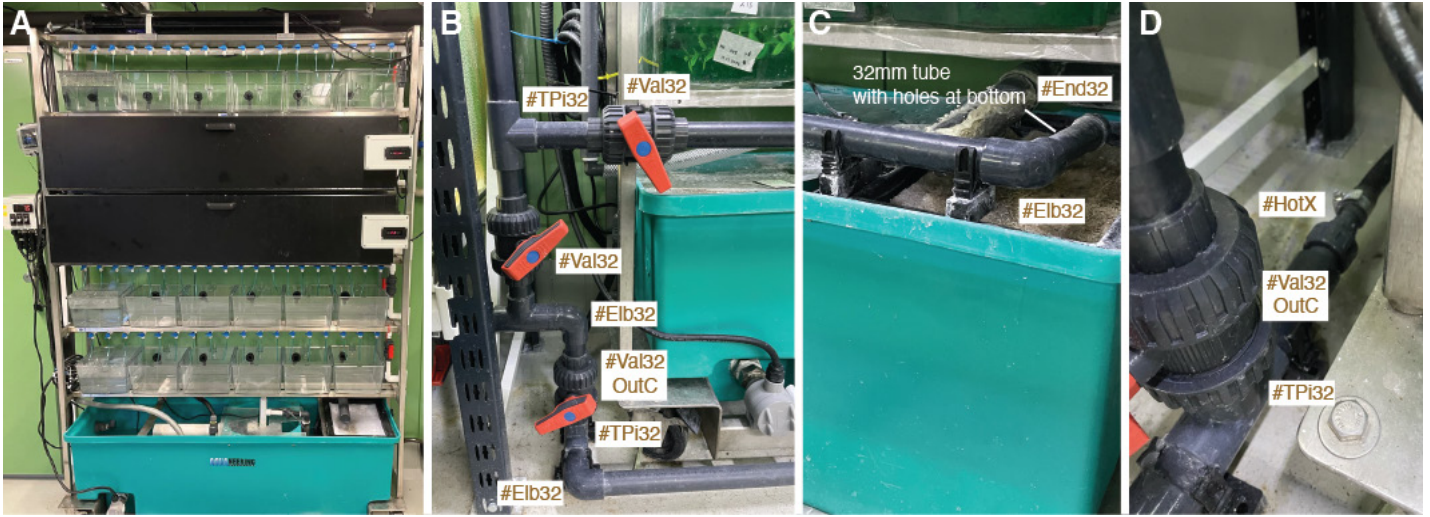

**Figure 8 | Integration of a regular zebrafish rack. A-D.** Zebrafish racks (A) are often used to raise young fish larvae. We integrated an available rack in our facility by adding it to outflow of the aquaria (B). A fraction of the water during water exchange goes into the zebrafish rack (C). Old water of the zebrafish rack is flowing hereby into the pipes of the water change system (D)

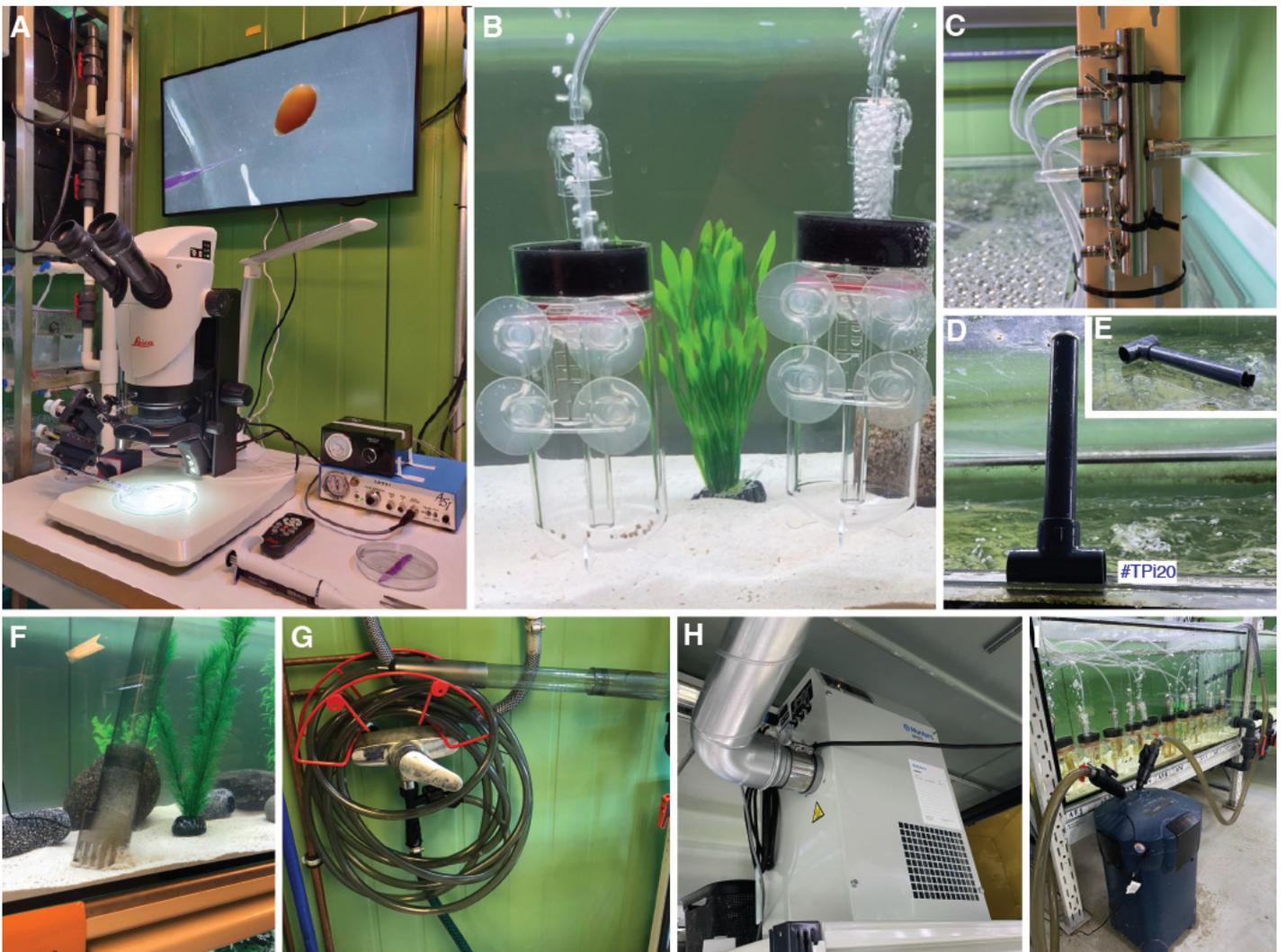

**Figure 9 | Additional items in the facility. (A)** A stereomicroscope with attached LCD screen and microinjection setup. **(B)** Egg tumbler for raising cichlid eggs. **(C)** Manifolds make managing pressured air for sponge filters and tumblers easier. **(D, E)** Self-constructed lid holders make cleaning easier. **(F, G)** A good investment is a cleaner that uses water pressure to generate suction (JBL Proclean Aqua In-Out Water Change Set). **(H)** In some facilities a dehumidifier might be needed to reduce the humidity that can quickly rise in a room with many aquaria. **(I)** An additional external filter might help to keep critical aquaria (e.g., for egg incubation) clean.

The lids are ideally already ordered in two pieces. One hole is added to the small piece for water inflow, while another hole is added to the larger piece to allow for easy opening (this also serves as a convenient way to feed the fish later (**Figure 4A, B**). On the next day, the first salt is added (necessary in our case). The aquaria need to run for four weeks before the first fish can be introduced. During this time, water changes should be monitored and tested. Temperature and parameters should be checked to confirm that they are within the correct range. Lastly, one must decide on fish food for differently sized fish, as well as fish fry and juveniles. Already during the initial cycling fish food should be added from time to time to allow the buildup of ammonia- and nitrite-oxidizing bacteria.

**Adding fish to the aquaria.** After a month, a few fish can be added to one of the aquaria. If they appear healthy after a few days, more can be gradually introduced. In our case the fish were evidently happy as they showed no signs of discomfort and even started to breed after a few days. Therefore, we also acquired ten cichlid egg tumblers and added them all to one aquarium to be able to raise eggs outside of the mouthbrooding females. Moreover, we added catfish (*Ancistrus spp.*) to all aquaria to reduce algae growth on the glass. During this period there might be fluctuation of water parameters and algae growth, so close monitoring is important.

**Learning from what is not working and further optimization.** While there were no major issues in this phase, there were a few things we realized that needed optimization. The biggest problem concerned the fact that the drainage system could not take that much water causing a mild flooding (that however was no issue in this waterproof room). To circumvent this problem, we added another valve to the outflow that limited the waterflow to just so much that the drainage could take it (**Figure 3B**). The second problem was some mild algae growth that we managed to get under to control by reducing the light intensity of the LED lights, while maintaining physiologically relevant light conditions. As this was a particular problem in the aquaria with the egg tumblers, we added an external filtration unit with an UV filter to this aquarium. Third, we realized later during summer month that the humidity in the room would increase quite drastically, why we decided to install an air dehumidifier that kept humidity around 50%.

## 4.2 | Expanding the facility.

**Adding more shelves and aquaria.** Expanding the facility should be already much easier than the first step, as there is confidence that the design works in practice. Challenges might only come by certain equipment not being available any more, which happened in our case for some of the lights and valves. Also costs for some items might increase, as it was the case for us for the aquaria. It might be therefore beneficial to expand the facility as soon as possible after the first phase has been completed successfully. Something else that is important to consider, is how much space is needed for safely working in the facility during daily work as well as maintenance and how much space is needed for additional task that are conducted in the room, while at this stage making a clear room plan is of great importance (**Figure 7**).

**Incorporating an existing zebrafish rack in the system.** A typical situation to be confronted with is if and how to incorporate existing material and premanufactured units into the system. In our case we inherited a fully functional zebrafish rack. These racks are expensive (one usually costs around 15.000€) and are excellent for keeping individuals separated during genotyping or for raising juveniles. Our challenge was that this rack needed water with the same parameters then the rest of the facility. Therefore, we decided, to reroute some of the water released during water changes into the zebrafish rack, the excess water is then released over the overflow of the rack (**Figure 8**). While this is no fresh water, we expected and could confirm that the water parameters (e.g., nitrate and nitrite) are excellent in the zebrafish rack, due to the regular water changes in the facility that we conduct and the amount of water that will be always replaced from the zebrafish rack (flowthrough of ca. 500 liter). This was a nice way to make sure that the water parameters of all systems align. It has a risk of infections spreading from the main aquaria to the zebrafish rack, but as aquaria with sick fish can be blocked via the valves these effects can be at least reduced. Also, the zebrafish rack has an additional UV filter next to mechanical and biological filtration that can further mitigate the risks. Still, it might be advised to not use such a rack for important fish.

**Adding working places for additional tasks.** Additional tasks that might be conducted in the fish facility are observations and manipulations on a microscope, microinjections for transgenesis and Crispr-Cas9, outreach events with demonstrations, e.g., of embryos at higher magnifications, but also storage space (and means to reach it) and cleaning equipment. At the same time, one wants to probably maximize the space for aquaria. We decided to add one small table, with a microscope and a microinjector (**Figure 9**). To aid training and outreach events, we obtained a microscope with an external monitor mounted to it as well as the possibility to record videos and images on an SD card. For storage space we use a shelf above the sink as well as the space on top of the aquarium shelves. To reach this space easily and organize it we bought a 3-step-ladder and several 40-liter wash baskets to conveniently store additional equipment.

## 4.3 | Establishing maintenance workflows

**Feeding.** The choice of food is highly important, as it can significantly impact water quality. We utilize high-quality granule food. This food is available in three sizes (<0.5 mm, 0.5-1 mm, 1-2 mm). We employ a color-coding system for our labeled aquaria to ensure the right-sized food is provided. For fry, we directly feed decapsulated brine shrimp eggs. Feeding eggs directly offers an advantage as they do not need to be hatched. They can be conveniently frozen in aliquots at -20°C. Before feeding, they need to be placed in a small cup with aquarium water for 15 minutes. As the eggs are stored in highly saline water, they expand when introduced into the less salty aquarium water. This expansion before ingestion is crucial, as ingestion of brine shrimp eggs before expansion could be fatal. Despite the higher cost of decapsulated eggs, this method significantly reduces the egg usage during hatching, preventing overproduction when only a few larvae need to be fed. This method also saves time, space, and mitigates the risk of forgetting to set up artemia.

**Daily Maintenance.** Ensuring the facility's functionality and the well-being of the fish demands daily checkups. This can be seamlessly integrated with feeding routines. Clear protocols and training are vital to identify sick or injured fish and know what actions to take in such cases. Filters, lights, temperature, water levels, leakage (water on the ground), water flow in zebrafish rack systems, humidity, and any signs of animal or technical issues should be diligently checked. These findings should be reported through a designated system, whether addressed or not, allowing others to review and providing an avenue for feedback. This ensures that relevant parties are informed of the situation. A report sheet might even be required by certain authorities. Daily maintenance usually takes 15-30 minutes, though it may take more time if issues need to be addressed.

**Weekly Maintenance.** Efficient water changes and maintenance are essential for conserving time and resources, while simultaneously upholding the facility's functionality and well-being of the fish. For our facility, which consists of 42 x 230-liter aquaria and a zebrafish rack, we conduct a weekly water change of 20%. This amount may seem substantial, but a two-week interval is too infrequent, and a non-fixed schedule is more challenging to organize. However, this weekly schedule ensures good water quality for heavily stocked aquaria and the zebrafish rack. During the water change, water is first released by opening the outflow valves. Emptying does not require close monitoring due to the built-in water level control. Subsequently, the outflow valves are closed, and the inflow valves are opened. While refilling the aquaria, salt is added to reach the desired hardness and pH-levels. For effective dissolution, the salt can be pre-dissolved in bottles placed in each aquarium, preventing oversights, and enabling quicker dissolution. While continuous presence is not necessary, a timer is essential to prevent overflow. It is vital to check everything is working properly one hour and one day after the water change, including water levels, temperature, and fish behavior. It is important to pay special attention to water temperature, as colder water is often used (as warm water can affect fish well-being as copper id often used for warm water pipes). Temperature dropping not excessively (below 20°C) is acceptable and may even stimulate mating. However, it's essential that the temperature swiftly returns to the normal range. Avoid conducting water changes before holidays or weekends, as issues are less likely to be detected during those times. Cleaning is done bi-weekly, involving vacuuming sand debris, and washing sponge filters in a bucket with aquarium water. A tap/faucet-based system is recommended for vacuuming, using tap water flow pressure to create suction that directly draws debris into the drainage (**Figure 9F, G**). Cleaning can be at the same time as the water release. The combination of cleaning and water change takes approximately 1½ to 2 hours, and having two people around can expedite the process, reduce errors, and make it more manageable. The time remains the same for water changes without cleaning, but there is a more extended break during water release and refill.

**Table 1:** Costs of establishing a fictional facility with 10 aquaria. For most items it is advised to buy one or more extra items in case they break (depending on storage space, how likely it is and what the collateral damage would be). Total cost for this setup with 10 aquaria without optional items would be around 9000€. *Costs are roughly given for a facility with 10 aquaria.

| Items | Total cost* |
|---|---|
| *Shelves and Aquaria* | |
| 10 x Heavy duty shelves (2 units, one for 6 and one for 4 aquaria) | ~3000€ |
| 10 x Aquarium 99 x 50 x 47cm, 8mm wall with drilled hole (Pavlica, Czech Republic) | ~1100€ |
| 1 x Delivery of aquaria (Price of one palette within Europe, ~450€) | ~900€ |
| 10 x Two-part acryl sheet cover and dividers | ~800€ |
| 40 x Suction Cup Divider Supports (e.g., Atyhao) | ~20€ |
| 10 x Camping mats to put under aquaria | ~50€ |
| *Water inflow system* | |
| 14 x PVC ball valves with adhesive sockets (Ø 20mm, Cepex) #Val20 | ~130€ |
| 8x coupling (screwable) with adhesive sockets (Ø 20mm) #Cou20 | ~10€ |
| 10x T-piece with adhesive sockets (Ø 20mm) #Tpi20 | ~4€ |
| 6x PVC 90° elbow piece with adhesive sockets (Ø 20mm) #Elb20 | ~2€ |
| 20x PVC 1m pipes (Ø 20mm) | ~20€ |
| 20x Pipe clamps (Ø 20mm) | ~6€ |
| Optional depending on water supply: 1x Ball valve (variable size and connection) #ValX | <10€ |
| Optional depending on water supply: 1x Hose #HosX | <10€ |
| Optional depending on water supply: 1x hose tail with adhesive socket (Ø 20mm) #HotX | <10€ |
| *Water drainage system* | |
| 14 x PVC ball valves with adhesive sockets (Ø 32mm, e.g., Cepex) #Val32 | ~190€ |
| 10 x PVC aquarium feed through with adhesive sockets (Ø 32mm) and 40mm outer thread (to fix feed through). Compatibility with pre-drilled hole has to be checked. | ~50€ |
| 8x coupling (screwable) with adhesive socket (Ø 32mm) #Cou32 | ~16€ |
| 10x T-piece with adhesive sockets (Ø 32mm) #TPi32 | ~7€ |
| 23x 90° elbow piece with adhesive sockets (Ø 32mm) #Elb32 | ~14€ |
| 10x PVC end piece with adhesive socket (Ø 32mm) #End32 | ~5€ |
| 10x pipe transition nipple with outer thread (1") and adhesive socket (Ø 32mm) #Nip32 | ~7€ |
| 10x 90° elbow piece with inner thread (1") and adhesive socket (Ø 32mm) #Elb32IT | ~10€ |
| 25x PVC 1m pipes (Ø 32mm) | ~42€ |
| 25x Pipe clamps (Ø 32mm) | ~9€ |
| 1x Pipe reduction ring with adhesive sockets (Ø 20mm x Ø 32mm) #Red2032 | ~1€ |
| *Construction tools* | |
| Pipe cutter | ~15€ |
| Pipe deburring tool | ~10€ |
| Cable ties (various sizes) | ~10€ |
| PVC cleaner (e.g., Tangit, 2x125ml) | ~15€ |
| PVC adhesive (e.g., Tangit, 500ml) | ~20€ |
| Aquarium glue (e.g., Hobby fix) | ~20€ |
| Cordless drill/driver with attachments and 20mm progressor hole saw (e.g., Bosch) | ~100€ |
| 1x high-density polyethylene block | ~80€ |
| Aquarium equipment | |

| Items | Total cost* |
|---|---|
| 10 x Sponge filters L (31 x 24 x 14 cm, e.g., from Senezal) | ~140€ |
| 10 x Aquarium lights (e.g., NICREW ClassicLED G2; 90-115cm, 32W) | ~500€ |
| 10 x Aquarium heaters (e.g., Sera Aquarium Heater Thermostat 150W for 75-200l) | ~200€ |
| 30 x Artificial Plants (3 per aquarium, e.g., Hobby Hygrophila) | ~270€ |
| 30 x Artificial hiding spots (3 per aquarium, e.g., Hobby marble cave) | ~180€ |
| 20 x Terracotta flowerpots (as hiding spots; should be cut in half) | ~100€ |
| 10 x Aquarium sand (expensive solution: Calcean African cichlid; 9kg per aquarium) | ~150€ |
| 10 x U-Shaped 8mm hanging glass thermometer (e.g., gototop) | ~100€ |
| 10 x Cichlid Egg tumblers (e.g., Ziss Aqua Egg-tumbler ZET-65) | ~320€ |
| 10 x 4 mm Silicone tubing for air distributors (e.g., JBL 2.5m, 4 mm inner Ø /6 mm outer Ø) | ~40€ |
| 5 x Silicone tubing for air distributors (8 inner Ø /10 mm outer Ø) | ~20€ |
| 5 x Air Distributors (e.g., from Psopp, 6-way or 10-way, input 8mm inner Ø, output 4 mm Ø) | ~75€ |
| 5 x Water-protected 4-way distribution sockets | ~100€ |
| 1 x JBL Proclean Aqua In-Out Water Change Set | ~45€ |
| *Additional Costs* | |
| Optional: Dehumidifying system (e.g., Munters) | ~4500€ |
| Optional: Stereomicroscope (Leica S9i StereoZoom) | ~6500€ |
| Optional: Screen and wall mount | ~500€ |
| Optional: Microinjection setup | ~2300€ |

**Table 2:** Running costs for a fictional facility (excluding rent, water, and electricity). *Costs are roughly given for one year for a facility with 10 aquaria. For some items (fine and medium food, artemia, less than the given amount is needed, while the total cost was reduced slightly).

| Item | Cost per item | Total Costs* |
|---|---|---|
| *Food* | | |
| Fine Food (<0.5mm; Nutramare) | 25€ per kg | 25€ |
| Medium Food (0.5-1mm; Nutramare) | 25€ per kg | 25€ |
| Large Food (1-2mm; Nutramare) | 25€ per kg | 25€ |
| Decapsulated Artemia Eggs (ArtemiaVita, 250ml = 40 million eggs) | 30€ for 250ml | 30€ |
| *Salt* | | |
| Epsom Salt (Magnesium Sulphate, $MgSO_4 \times 7H_2O$) | 24€ per 5kg | 24€ |
| Natron Salt (Sodium hydrogen carbonate $NaHCO_3$) | 14€ per 5kg | 28€ |
| *Water parameter measurement* | | |
| Strips to check water parameters (e.g., Tetra Test 6-in-1) | 9€ | 9€ |

**Monthly and yearly maintenance.** Regular, more thorough checks are recommended. Monthly checks should encompass checking essential water parameters, including Nitrate and Nitrite levels. Additionally, parameters such as pH, hardness, chlorine, chloramine, copper, and heavy metals should be also checked but depend on the quality of the water used for refilling aquaria. Some algae growth is tolerable, but if it becomes problematic because of the type or amount of algae, adjustments in light levels, using an external filter with UV, performing more extensive water changes, adding more algae-eating catfish, or even resorting to chemical treatments can help control it. Keeping track of fish and knowing when to breed them to maintain stocks can become challenging. Conducting an inventory once or twice a year can help reduce chaos without being overly time-consuming. It also aids in fair space distribution and accommodating new projects.

**Establishing safety measures and contingency plans.** Even when everything seems well-controlled, unexpected events can occur. Aquaria might break, equipment can malfunction, power outages, heatwaves, or disease outbreaks can happen. It is prudent to have a written contingency plan outlining the steps to take when something goes wrong. This is especially critical in severe situations where quick action is necessary. Furthermore, maintaining backups of filters, heaters, external filters, and air pumps provides an alternative means of oxygenation.

## 5. | COSTS

### 5.1 | Costs of building a facility.

Costs of building a facility (**Table 1**) will be influenced by many things including the space, needs, expertise and time as well as where you are placed. Still, as a reference we wanted to provide a summary of the approximate costs per item and a facility with 10 aquaria to allow easy calculation.

### 5.2 | Costs of running a facility.

For the cost of running a facility (**Table 2**) there will be a wide range of items needed that depend on the specific needs. Also here, we provide costs for a fictional facility with 10 aquaria to allow easy calculation.

## 6. | CONCLUSION

In summary, we have detailed the construction of a purpose-built facility designed specifically for tropical fish. We trust that our thought processes and decision-making can offer valuable insights to researchers who are contemplating the establishment of their own facilities. While variations might exist from one location to another, having a tangible example of how it can be achieved can provide guidance and complementary advice, enriching one's personal experience. Having undertaken this endeavor ourselves, despite occasional moments of stress, proved to be an immensely rewarding and enjoyable process. At present, our facility accommodates 15 distinct species, housing over 1000 fish, with nearly all species engaging in regular breeding activities. The simplicity of maintenance translates to significant time savings, while still having granted us the luxury of customizing our setup according to our preferences.

## ACKNOWLEDGMENTS

CFK would like to thank Ralf Schneider for his invaluable expertise in aquaria design and construction, as well as to Jan Gerwin and Ralf Schneider for their comprehensive teachings on fish keeping and maintenance. CFK is grateful for the exceptional advice and thoughts about fishes and fish facilities provided by (sorted in alphabetical order) Karoliina Alm, Ehsan Pashay Ahi, Ingo Braasch, Ulrika Candolin, Nicolas Kolm, Axel Meyer, Nikolai Piavchenko, Craig Primmer, Emília Santos, Walter Salzburger, and Jukka-Pekka Verta. We also extend our thanks to the Association of Finnish Cichlid Hobbyist "Ciklidistit ry" (https://www.ciklidistit.fi), particularly Jari Nyman, for their engaging discussions and a recent aquarium donation. Importantly, we also want to express our gratitude to the Institute of Biotechnology, HiLIFE for their financial support of our facility and the Faculty for Biological Environmental Sciences of the University of Helsinki for the space, both of which has been crucial to the realizations of the here presented facility and our research. CFK acknowledges the use of AI language models, specifically ChatGPT, for proofreading and improving the writing of this manuscript as well as the use of Adobe Illustrator software for creating and refining the figures used in this publication and Adobe InDesign to layout the manuscript.

## CONFLICTS OF INTEREST

The authors declare that there are no conflicts of interest.

## AUTHOR CONTRIBUTIONS

CFK designed the facility and wrote the manuscript. MK, JH, NK were incremental at different stages of the process to make the facility a working facility from helping to build it (MK) to establishing, improving, and maintaining workflows (MK, JH, NK). Everybody read and commented on the manuscript.

## DATA AVAILABILITY STATEMENT

All data (i.e., engineering data, cost estimates, and biological data about water parameters) are directly presented in the manuscript and its accompanying tables.

## REFERENCES

Carruthers, Steven. 2015. "Small-Scale Aquaponic Food Production." Practical Hydroponics and Greenhouses, no. 152: 42–46. https://search.informit.org/doi/10.3316/informit.903224464910191.

Magalhaes, Isabel Santos, and Antonia Geraldine Patricia Ford. 2022. "The Amazing Diversity of Cichlid Fishes." Frontiers for Young Minds 10: 544098. https://doi.org/10.3389/frym.2022.544098.

McNabb, Adrian, Kirsty Scott, Elke von Ochsenstein, Kirsten Seufert, and Matthias Carl. 2012. "Don't Be Afraid to Set Up Your Fish Facility." Zebrafish 9 (3): 120–25. https://doi.org/10.1089/zeb.2012.0768.

Paige, Candler, Bailey Hill, Joseph Canterbury, Sarah Sweitzer, and E Alfonso Romero-Sandoval. 2014. "Construction of an Affordable and Easy-to-Build Zebrafish Facility." Journal of Visualized Experiments, no. 93: e51989. https://doi.org/10.3791/51989.

Rakocy, J.E. 2006. "CICHLIDAE. Cultured Aquatic Species Information Programme." FAO 2023, Fisheries and Aquaculture Division [Online]. 2006. https://www.fao.org/fishery/en/culturedspecies/oreochromis_niloticus/en.

Santos, M. Emília, João F. Lopes, and Claudius F. Kratochwil. 2023. "East African Cichlid Fishes." EvoDevo 14 (1): 1. https://doi.org/10.1186/s13227-022-00205-5.